\documentclass[preprint,showpacs,preprintnumbers,amsmath,amssymb]{revtex4}
\usepackage{graphicx,epstopdf}

\usepackage{dcolumn}
\usepackage{bm}
\usepackage{amssymb}
\begin{document}
\title{Observation of low-lying excitations of electrons in coupled quantum dots}
\author{C\'{e}sar Pascual Garc\'{i}a, Sokratis Kalliakos, Vittorio Pellegrini}
\affiliation{NEST CNR-INFM and Scuola Normale Superiore, Piazza dei
Cavalieri 7, I-56126 Pisa, Italy}
\author{Aron Pinczuk*, Brian S. Dennis, Loren N. Pfeiffer, Ken W. West}
\affiliation{Bell Labs, Lucent Technologies, Murray Hill, New
Jersey}
\affiliation{*Dept of Physics, Dept of Appl.~Phys.~and Appl.~Math.,
Columbia University, New York, New York}
\date{\today}
\begin{abstract}
Tunneling excitations of electrons in dry-etched modulation-doped
AlGaAs/GaAs coupled quantum dots (QDs) are probed by resonant
inelastic light scattering. A sequence of intra- and inter-shell
excitations are found at energies determined by the interplay
between the QD confinement energy $\hbar \omega _o$ and the
tunneling gap $\Delta _{SAS}$, the splitting between the symmetric
and anti-symmetric delocalized single particle molecular states.
The narrow line-widths displayed by electronic excitations in these
nanostructures indicate promising venues for the spectroscopic
investigation of entanglement of electron states in these
artificial molecules.
\end{abstract}
\maketitle
\par
Semiconductor quantum dots (QD's) are regarded as artificial atoms
with electron properties that can be tailored on
demand~\cite{Reimann}. In this framework, artificial molecules can
be realized with two interacting quantum dots~\cite{Vanderwiel}.
It has been proposed that coupled quantum dots are of particular
relevance for quantum computation schemes~\cite{DiVincenzo,
loss98}. In fact, the tunneling gap $\Delta _{SAS}$ that splits
the QD levels into their symmetric and anti-symmetric combinations
can modulate the exchange interaction and provides a route towards
entanglement of two spins~\cite{Burkard}. Electron states in
coupled double QD are expected to be less sensitive to dephasing
mechanisms linked to coupling of spin and charge states that occur
in a single QD environment. Coupled QDs also offer ways to study
novel spin and charge collective phases and quantum phase
transitions at the nanoscale~\cite{Rontfases}.
\par
The remarkable physics and applications of coupled QDs is manifested
in several experimental studies. Transport experiments based on
tunneling  have extensively investigated inter-dot coupling
effects in the strong and weak coupling regimes
~\cite{ota,petta,tarucha97}. These measurements have provided
evidence for the impact of symmetric and anti-symmetric states in
the Coulomb blockade spectra and have demonstrated slower
relaxation rates for spin states in laterally coupled QDs and
their coherent manipulation. Inter-band luminescence experiments
also have been carried out in self-assembled InAs coupled
QDs~\cite{krenner}.
\par
In this letter we report the observation of low-lying neutral
excitations of electrons in vertically coupled GaAs/AlGaAs double
QDs nanofabricated by e-beam lithography and dry etching. Resonant
inelastic light scattering spectra access excitations linked to
the tunneling and quantum confinement degrees of freedom of
electrons. The spectra reveal sharp excitations, with FWHM below
200$\mu eV$, that demonstrates the high quality of these etched
nanostructures. These QD systems could be suitable for exploration
of entanglement of electron states that enter in quantum
computation schemes based on the optical manipulation of spin and
charge.
\par
Eigenstates of a parabolic  QD in the non-interacting scheme are
described by the Fock-Darwin (FD) levels. In vertical QDs the coupling
between the two dots leads to the splitting of the single dot
levels in bonding (symmetric) and antibonding (antisymmetric)
levels separated by the tunneling gap $\Delta_{SAS}$. The single
particle eigen-energies of parabolic double QDs at zero magnetic
field can thus be modeled by the following equation:
\begin{equation}
\label{2DCQD}
E_{N,P}= \hbar \omega_0 (2n + |m| + 1)+ P \Delta_{SAS} = \hbar \omega_0 (N+ 1)+ P \Delta_{SAS},
\end{equation}
where $n=0,1,\ldots$, $m=0\pm1, \ldots$ are the radial and
azimuthal quantum numbers, respectively, $N =2n+|m|$ identifies
shells with well-defined atomic-like parities, $\hbar\omega_0$ is
the confinement energy and $P$ is the extra degree of freedom
labeled as a pseudospin. $P$  takes values -1/2 for symmetric
levels and 1/2 for antisymmetric levels.  This peculiar
energy-level structure yields an excitation spectrum characterized
by tunneling or pseudospin modes constructed from intra- or
inter-shell excitations of electrons from the symmetric to the
anti-symmetric QD levels. Figure 2a shows the FD sequence
following Eq.1 in the case in which $\Delta _{SAS}$ is slightly
larger than $\hbar \omega _0$.
\par
Parity selection rules applied to parabolic double coupled QDs
establish that monopole transitions with $\Delta m=0$ ($\Delta N =
0,2,4,\ldots$) and $\Delta P = 0,\pm 1$ are the strongest
intensity modes active in light scattering experiments in a
backscattering geometry \cite{schuller99,cesar05}. The
single-particle representations of intra- and inter-shell
pseudospin excitations associated with changes of $N$ and $P$ are
shown in Fig.2a as vertical arrows.  In agreement with this scheme
we found in our experiments two lowest-energy peaks in the
low-temperature inelastic light scattering spectra at energies
below 1 meV that were assigned to the intra-shell $\Delta P$ =1
excitation at  $\Delta _{SAS}$ and to the inter-shell pseudospin
mode at $2\hbar\omega _{o}-\Delta _{SAS}$. These peaks  are
remarkably sharp with a full width half maximum (FWHM) of 200 $\mu eV$
and allow direct measurement of $\Delta _{SAS} = 1 \pm 0.1meV$ and
$\hbar \omega _{0} = 0.8 \pm 0.05 meV$ in our system. Consistent
with this assignment we found a broader inter-shell monopole mode
at $2\hbar \omega _{o} = 1.8 \pm 0.1meV$. Contrary to the case of
single QDs,  this inter-shell excitation has a marked temperature
dependence and disappears around 15K. The measured
activation gap is consistent with the value of $2\hbar \omega
_{o}-\Delta _{SAS}$, offering further evidence of the impact of
inter-dot coupling on the excitation spectrum. Finally an
additional broad mode is observed around $2.4 \pm 0.4 meV$
corresponding to $4\hbar \omega_{0}-\Delta _{SAS}$.
\par
The sample was fabricated starting from an 18 nm wide, symmetrically modulation-doped Al$_{0.1}$Ga$_{0.9}$As/GaAs double quantum well with 6nm Al$_{0.1}$Ga$_{0.9}$As barrier separation (see Fig.1). The tunneling gap measured by light scattering in the double layers (data not shown) is 0.8 meV. The low-temperature electron density and mobility measured from Subnikov De Haas are $n_e=3\times 10^{11}$ cm$^{-2}$ and $2.7\times 10^6$ cm$^2$/Vs, respectively. The lateral confinement was produced by inductive coupled plasma reactive ion etching (ICP-RIE). To this end a 30 nm thick nickel mask was first deposited on top of the sample. Coupled QD arrays (with sizes $100 \mu m \times 100 \mu m$ containing $10^4$ single coupled QD replica) with lateral dot diameters of about $400$ nm were defined by electron beam lithography. This value of lateral diameter was chosen to yield a confinement energy of the order of the tunneling gap and a large electron occupation (above 100). In this many-electron case, in fact, the light scattering response at resonance is dominated by single-particle transitions linked to Eq.1 and not corrected by dynamical many-body effects \cite{schuller99}. Deep etching (below the doping layer) was then achieved by using a mixture of BCl$_{3}$/Cl$_{2}$/Ar in the ICP-RIE and applying low voltages. The nickel mask was removed before the optical studies. Top panels in Fig. 1 show scanning electron microscope (SEM) images of the coupled dots in the array. Light scattering experiments were performed in a backscattering configuration ($q\le 2\times 10{^4}$ cm$^{-1}$ where $q$ is the wave-vector transferred into the lateral dimension) with temperatures down to $T=1.9$ K. A tunable ring-etalon Ti:sapphire laser was used and the scattered light was collected into a triple grating spectrometer equipped with a CCD detector.
\par
\begin{figure}
\includegraphics[width=7cm]{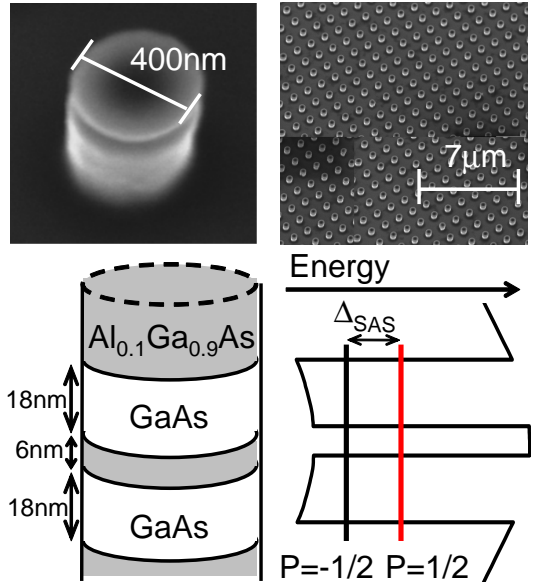}
\caption{\label{fig1} Top: SEM images of one dry-etched quantum dot (QD) and the $100\mu m$ x $100\mu m$ array composed by $10^{4}$ coupled QDs separated by $1 \mu m$. Bottom: Schematic description of the vertical double QD and the corresponding energy diagram in the conduction band. The two lowest symmetric and antisymmetric levels are also shown. $P$ is the pseudospin, $\Delta _{SAS}$ is the tunneling gap of the coupled QD.}
\end{figure}
\par
\begin{figure}
\includegraphics[width=8cm]{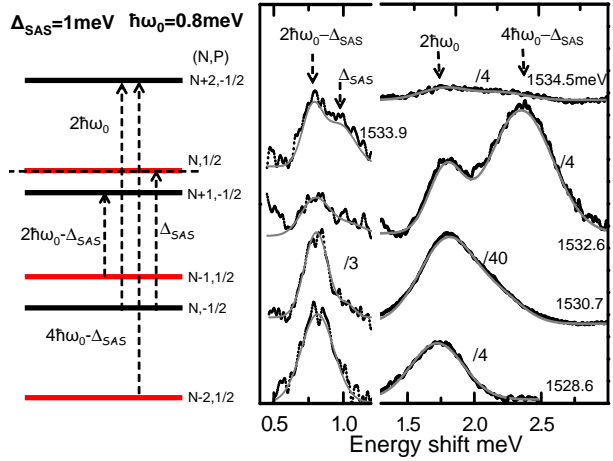}
\caption{\label{fig2} Left:  schematic representation of energy levels and transitions in the coupled quantum dot. $N$ and $P$ are the shell and pseudospin quantum numbers, respectively. Black and red lines represent symmetric and antisymmetric levels, respectively. The dotted line marks the position of the Fermi level that accounts for the observed reduced intensities of the two lowest-energy modes. Right panel: Resonant inelastic light scattering spectra at 1.9K and at different laser energies (shown in the figure) in depolarized configuration. Intensities were scaled by factors indicated in the figure.}
\end{figure}
\par
Figure 2 shows the resonant inelastic light scattering spectra at different excitation energies and T=1.9K after conventional subtraction of the background due to interband luminescence.  The laser with intensity of $0.1 W/cm^2$ was tuned between 1528.6meV (bottom spectrum) to 1534.5meV (top spectrum) to explore different resonances. The spectra were taken in a depolarized configuration with perpendicular polarizations of the incoming and outgoing light in order to reduce the stray laser light.
\par
Spectra shown in Fig.2 are remarkably different from those found in single QDs  with the same lateral diameter ~\cite{physicaE}. In single QDs the resonant inleastic light scattering spectra display a sequence of peaks equally spaced in energy by $2\hbar \omega _{0}$ with a FWHM of 1 meV in agreement with data obtained by other groups~\cite{schuller99}. The spectra of double QDs are instead characterized by two sharp (FWHM = 0.2 meV) low-energy peaks at $2\hbar \omega _{0}-\Delta_{SAS}$ and $\Delta _{SAS}$. The peak at 1.8meV corresponding to  $2\hbar\omega_0$ is thus assigned to the conventional inter-shell $\Delta N$=2 mode also observed in the single QDs. The peculiar energy level structure of the coupled QDs is additionally revealed by the fourth highest-energy peak shown in Fig.2 and observed at an energy corresponding to $4\hbar\omega_0 -\Delta_{SAS}$ = 2.4 meV. The energy of this peak shifts from 2.2 to 2.6 meV depending on the laser excitation wavelength. This behavior can be linked to nonparabolicity effects whose impact increase with the energy of the mode and to partial overlap with the $2\hbar\omega_0+\Delta_{SAS}$ inter-shell pseudospin mode expected at $\sim$2.8 meV. More excitations were detected at higher energies (data not shown) with decreasing intensity and with broader signal corresponding to higher inter-shell excitations combined with the symmetric and antisymmetric states. It can be noted that the energy of the intra-shell pseudospin mode ($\Delta _{SAS}$ = 1 meV) is higher than the the single-particle excitation energy at the bare tunneling gap measured in the double quantum well prior to nanofabrication ($\sim$0.8 meV). This is probably due to partial depletion of electrons caused by the etching processes.
\par
The difference in the intensities between the modes below 1 meV and those at higher energies is remarkable. It suggests partial population of the two highest-energy occupied levels as indicated by the position of the Fermi level shown by the dotted line in the left part of Fig.2. These two levels are the excited states associated with the two sharp low-lying transitions. Their partial population explains the  reduced intensities of the two modes due to phase space filling effects. The results in Fig.2 therefore suggests that light scattering can be applied to determine both energies and population of molecular states in coupled QDs.
\par
\begin{figure}
\includegraphics[width=7.5cm]{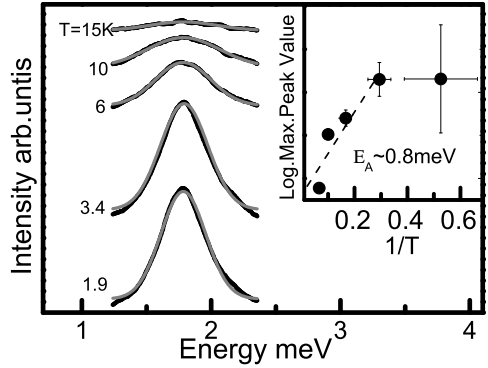}
\caption{\label{fig3} Temperature dependence of the monopole inter-shell transition at $2\hbar \omega _o$. Laser intensity and energy are 0.1 W/cm$^2$ and 1567 meV. Spectra are presented after conventional subtraction of background due to luminescence. The inset shows an Arrhenius plot of the integrated intensity with an activation energy of 0.8 meV.}
\end{figure}
\par
Further evidence of the impact of inter-dot coupling arises from the temperature behavior of the conventional intra-shell monopole excitation at $2\hbar\omega_0$. Contrary to the single quantum dot case, in fact, where this excitation remains unchanged up to temperatures above 30K, here a significant change of the signal intensity occurs at much lower temperatures and the inter-shell mode disappears around 15K with an activation energy of 0.8 meV as displayed in Fig.3. The activation gap is consistent with the value of $2\hbar\omega_0-\Delta_{SAS}$, the gap  separating the highest-energy occupied level (an antisymmetric state with shell number $N$) from the lowest-energy unoccupied level (a symmetric state with shell number $N+2$). This behavior thus  offers further evidence of the impact of inter-dot coupling in the excitation spectrum.
\par
In conclusion, we reported the first measurements of excitations
of electrons in nanofabricated vertically-coupled quantum dots.
The spectra reveal a low-lying intrashell pseudospin mode across the
tunneling gap as well as inter-shell excitations resulting from
the interplay between the confinement energy and the tunneling
gap. The results presented here suggest that, by offering access
to molecular-like excited states in the coupled QDs, the light
scattering methods can provide a wealth of quantitative
information on the energy level sequence, level occupation and
tunneling gap in double QDs. Further work in the few-electron
occupation regime should  address the interplay between spin and
pseudospin excitations in the coupled dots.
\par
We acknowledge support from the Italian Ministry of Foreign
Affairs, Italian Ministry of Research (FIRB-RBAU01ZEML), European
Community's Human Potential Programme (HPRN-CT-2002-00291),
National Science Foundation (DMR-03-52738), Department of Energy
(DE-AIO2-04ER46133). We are grateful to SENTECH-Berlin for
allowing us to use the ICP-RIE machine. We acknowledge useful
discussions with Massimo Rontani, Guido Goldoni and Elisa
Molinari.


\begin{thebibliography}{99}

\bibitem{Reimann}
{S.M. Reimann and M. Manninen, Rev. Mod. Phys. {\bf 74,} 1283 (2002).}

\bibitem{Vanderwiel}
{V.G. Van Der Wiel {\em et al.}, Rev. Mod. Phys. {\bf 75}, 1 (2003)}

\bibitem{DiVincenzo}
{D.P. DiVincenzo {\em et al.}, Nature {\bf 408}, 339 (2000)}

\bibitem{loss98}
{D. Loss and D. P. DiVincenzo, Phys Rev. A {\bf 57,} 120 (1998).}

\bibitem{Burkard}
{G. Burkard {\em et al.}, Phys. Rev. B {\bf 59}, 2070 (1999)}

\bibitem{Rontfases}
{M. Rontani {\em et al.}, Phys. Rev. B {\bf 69}, 085327 (2004)}

\bibitem{ota}
{T. Hatano {\em et al.}, Science {\bf 309} 268 (2005)}

\bibitem{petta}
{J.R. Petta {\em et al.}, Science {\bf 309} 2180 (2005)}

\bibitem{tarucha97}
{T.H. Wang, S. Tarucha, Appl. Phys. Lett. {\bf 71} 2499 (1997)}

\bibitem{krenner}
{H.J. Krenner {\em et al.}, Phys. Rev. Lett. {\bf 94}, 057402 (2005)}

\bibitem{schuller99}
{R. Strenz {\em et al.},  Phys. Rev. Lett. {\bf 73} 3022 (1995); D.J. Lockwood {\em et al.}, Phys. Rev. Lett. {\bf 77}, 354 (1996); C. Schuller {\em et al.} Phys. Rev. Lett. {\bf 80}, 2673 (1999)}

\bibitem{cesar05}
{C. Pascual Garc\'ia {\em et al.} to appear in Physical Review Letters (2005).}

\bibitem{physicaE}
{C. Pascual Garc\'ia {\em et al.} to appear in Physica E (2005).}


\end{thebibliography}
\end{document}